\documentclass[twocolumn,secnumarabic,amssymb, nobibnotes, aps, prd]{revtex4-1}
\usepackage{amsmath}
\usepackage{nicefrac}
\usepackage{graphicx}
\usepackage{wrapfig}
\usepackage{setspace}
\usepackage{hyperref}
\usepackage{url}

\newcommand{\ee}[1]{\begin{align}#1\end{align}}

\begin{document}

\title{Shape is Destiny: Configurational Entropy as a Lifetime Predictor and Pattern Discriminator for Oscillons}

\author{Marcelo Gleiser}
\email{Marcelo.Gleiser@dartmouth.edu}
\affiliation{Department of Physics and Astronomy\\ Dartmouth College,Hanover, NH 03755, USA}

\author{Michelle Stephens}
\email{Michelle.M.Stephens.GR@dartmouth.edu}
\affiliation{Department of Physics and Astronomy\\ Dartmouth College,Hanover, NH 03755, USA}

\author{Damian Sowinski}
\email{Damian.Sowinski@dartmouth.edu}
\affiliation{Department of Physics and Astronomy\\ Dartmouth College,Hanover, NH 03755, USA}

\date{\today}

\begin{abstract}

Oscillons are long-lived, spherically-symmetric, attractor scalar field configurations that emerge as certain field configurations evolve in time. 
It has been known for many years that there is a direct correlation between the initial configuration's shape and the resulting oscillon lifetime: a shape memory. 
In this paper, we use an information-entropic measure of spatial complexity known as {\it differential configurational entropy} (DCE) to obtain estimates of oscillon lifetimes in scalar field theories with symmetric and asymmetric double-well potentials.  
The time-dependent DCE is built from the Fourier transform of the two-point correlation function of the energy density of the scalar field configuration. 
We obtain a scaling law correlating oscillon lifetimes and measures obtained from its evolving DCE. 
For the symmetric double-well, for example, we show that we can apply DCE to predict an oscillon's lifetime with an average accuracy of $6\%$ or better. 
We also show that the DCE acts as a pattern discriminator, able to distinguish initial configurations that evolve into long-lived oscillons from other nonperturbative short-lived fluctuations.
 
\end{abstract}

\maketitle

\section{Introduction}

\noindent
From one-dimensional kinks and sine-Gordon solitons \cite{Rajaraman} to spherically-symmetric Q-balls \cite{Coleman1} and bounce solutions in false vacuum decay \cite{Coleman2, Coleman3}, self-interacting scalar fields are known to produce a variety of static spatially-bound configurations, with properties that depend on the number of spatial dimensions and the details of the interactions. 
The stability (instability) of such configurations is linked to the existence (nonexistence) of a conserved charge due to an underlying symmetry of the Lagrangian describing the field and its interactions. 
For example, in one spatial dimension, $\phi^4$ kinks owe their stability to the discrete ${\cal Z}_2$ symmetry of the vacuum manifold, while the simplest Q-balls to a globally-conserved $U(1)$ charge. 

For models lacking a symmetry as, for example, in scalar field theories with asymmetric effective potentials, bounce configurations describe the decay of a metastable or false-vacuum state. 
As is well-known, in $d$ spatial dimensions the lifetime of such a state can be computed semi-classically using the saddle-point approximation and is given by $\tau \sim \exp[S_E(\phi_b)/\hbar]$, where $S_E(\phi_b)$ is the $(d+1)$-dimensional Euclidean action of the bounce (or bubble) configuration $\phi_b$. 
As long as $S_E(\phi_b)/\hbar \gg 1$, the semi-classical approximation can be trusted. 
Otherwise, it may be necessary to add nonperturbative corrections to the decay rate \cite{GleiserHeckler}. 
For finite-temperature field theories \cite{LeBelac}, one uses a compact time dimension to obtain the lifetime $\tau \sim \exp[S_T(\phi_b)/k_BT]$, where $S_T(\phi_b)$ is the $d$-dimensional action with finite-temperature corrections added to the effective potential, $T$ is the temperature, and $k_B$ is Boltzmann's constant.

Another class of unstable field configurations is known as oscillons, long-lived oscillating solutions of a variety of scalar field theories. (For an incomplete list of references see \cite{Bogolubsky, Gleiser, GleiserSornborger, Fodor, Honda, Kasuya, GrahamStamatopoulos, Graham, GleiserSicilia, Farhi, Hertzberg, Salmi, Saffin, Mukaida}.) 
Over the past few decades, oscillons were shown to play important roles in the dynamics of phase transitions \cite{GleiserHowell, GleiserThorarinson}, in early universe inflationary dynamics \cite{Amin, GleiserGrahamStamatopoulos, AminEasther}, and as possible sources of gravitational waves \cite{Zhou, Antusch}, among other topics. 

An essential open question in oscillon models is their lifetime. 
Although progress has been made in computing classical \cite{GleiserSicilia, Mukaida} and quantum \cite{Hertzberg, SaffinTranberg} radiating rates, oscillon lifetimes have so far been estimated by solving the equation describing the time-evolution of the oscillon and watching it decay. 
Given their longevity, such methods are numerically costly, even with ingenuous conformal coordinate transformations as in Refs. \cite{Fodor, Honda}. 
It would thus be desirable to have a numerically-efficient method, capable of predicting oscillon lifetimes and of providing new insights into their complex dynamics. 
This is the main task of the present manuscript.

On a companion paper, we will revisit the question of false vacuum decay, focusing, as is the case here, on the lifetime of the metastable state. 
As we will show, the same tools we develop to address the question of oscillon lifetime can be used for vacuum decay, with results that correlate strongly with those obtained using the traditional decay rates computed with the saddle-point approximation.

In both studies, the unifying concept that brings together the longevity of oscillons and the decay of false vacua is information theory. 
In recent years, a new measure of spatial complexity known as configurational entropy (CE) was proposed \cite{GleiserStamatopoulos1} and applied to a wide variety of topics, including spontaneous symmetry breaking \cite{GleiserStamatopoulos2}, the stability of Q-balls \cite{GleiserSowinski1} and of neutron and boson stars \cite{GleiserJiang1}, brane-world models \cite{Correa}, glueballs \cite{Bernardini}, anti-de Sitter black holes \cite{Braga, Lee}, the determination of the critical point in phase transitions \cite{GleiserSowinski2}, and spontaneous emission in hydrogen atoms \cite{GleiserJiang2}. 
Inspired by Shannon's information entropy \cite{Shannon}, CE is constructed from the Fourier spectrum of spatially-localized or periodic field configurations and provides a measure of the spatial complexity of the configuration related to its localization: the more spatially-localized the configuration, the more spread-out it is in momentum space and the higher its CE. 
As the configuration begins to approach uniformity, complexity is lost, and CE begins to drop. 
In that sense, a single sinusoidal mode has zero CE (only one momentum mode), as does a discrete set of $N$ equiprobable modes. 
Ref. \cite{GleiserSowinski2} introduced a detailed informational interpretation of the CE, where the ``message'' is a particular configuration described by a spatially-bound or periodic function, and the ``alphabet'' is given by the momentum modes that compose the configuration with specific weights (probabilities), as obtained from its Fourier transform. 
This way, each field configuration -- or message -- will have a specific informational signature in momentum space with quantifiable complexity. 
Though the term CE has been used in prior work \cite{GleiserStamatopoulos1,GleiserStamatopoulos2,GleiserSowinski1,GleiserJiang1,GleiserSowinski2,GleiserJiang2}, this work wishes to distinguish the formal definition of CE, which applies only to discrete systems, and differential CE (DCE) which applies to continuous ones.

The interest in applying our information theory methodology to oscillons and false vacuum decay is twofold. 
First, and with direct relevance to the present work, as shown in applications to solitons \cite{GleiserSowinski1} and gravitationally-bound states (nonrelativistic and relativistic stars) \cite{GleiserJiang1}, the stability of a given configuration is closely tracked by its corresponding DCE, with the instability point given by the maximum of the DCE curve with respect to the parameter determining its physical properties. 
For example, for Q-balls it would be the pair ($\omega,~b$), where $\omega$ is the angular frequency of the complex scalar field and $b$ the parameter determining the shape of its effective potential, while for relativistic and boson stars it would be the star's central density $\rho_0$. 
Second, it is possible to adapt the definition of a quantity known in information theory as the Kullback-Leibler divergence \cite{KullbackLeibler} to field theory, as shown originally in Ref. \cite{GleiserStamatopoulos2} and, in cosmology in Ref. \cite{GleiserGraham}. 
Such a quantity is extremely useful in applications where one needs to pick out structure from a noisy background, be it quantum or thermal. 
Here, we will explore the first application above, applying the DCE to obtain an accurate estimate of oscillon lifetimes. 
We will also show how the DCE can be used as a pattern discriminator, picking out oscillons from other short-lived subcritical bubbles.

This paper is organized as follows: Section II briefly reviews the formalism for configurational entropy and considers how to apply it to either continuous or discrete systems. 
Section III reviews oscillons and presents our main results. 
Section IV concludes with a few remarks and future research. 

\section{Configurational Entropy}
Configurational Entropy is inspired by Shannon entropy, a measure of information that made its debut in 1948 \cite{Shannon}:
\ee{\label{ShannonEntropy}
S = -\sum_{a\in \mathcal A}p_a\log p_a.
} 
Here, $\mathcal{A} = \{a_1, a_2, ..., a_N\}$ is an alphabet, consisting of letters which appear with probability $p_i = p(a_i)$ in some corpus of messages.
Alongside entropy, the information content of a particular letter is defined as $I(a) = -\log_2 p(a)$.
The two are related by the fact that the former is the expected value of the latter:
\begin{equation}\label{ShannonInfo}
\langle I \rangle = \sum_{a\in\mathcal A} p(a)I(a) = -\sum_{a\in\mathcal A} p(a) \log_2 p(a) = S.
\end{equation}
Information content is the minimum number of bits needed to encode a letter to achieve a maximal transmission rate of messages between an information emitter and receiver, the {\it channel capacity}. 
An illuminating interpretation of entropy is the expected number of {\it yes-no} questions needed to reveal the identity of a randomly letter drawn given a corpus, the $\mathcal Q$-interpretation.

If we know nothing of the corpus from which a letter is drawn, we must ask $\log_2 N$ {\it yes-no} questions (in expectation) to reveal the symbol; equivalently, the number of bits needed to store the symbol is $\log_2 N$.  
If we know the corpus, then common letters require fewer than $\log_2 N$ bits to store, while rare letters could require much more.
For example, in the corpus of English literature the letter $e$ has the highest probability of occurrence at $p(e) \simeq 0.09$ and an information content of $I(e) \simeq 3.5$ bits, while the letter $q$ has $p(q) \simeq 0.01$ and $I(q) \simeq 10.3$ bits \cite{MacKay}. 
If we didn't know we were drawing letters from the English corpus, then each letter would have probability $1/26\simeq  .04$ and an information content of $4.7$ bits.
The entropy in the former case is $4.17$ bits, while in the latter it is $4.7$ bits \cite{Sowinski:English}.
Knowing the corpus from which a letter is drawn will (almost) always decrease the entropy of an alphabet relative to not knowing the corpus.
Ignorance manifests itself in a uniform distribution over possible outcomes, $p(a) = 1/N$, resulting in entropy being maximized at $\langle I \rangle = \log_2 N$. 
This has a direct impact on digital storage of images: the more random an image is the more space it will take up on your computer. 

CE is constructed on a finite system using the above ideas \cite{Sowinski:Thesis}.
What happens when we allow our alphabet to grow to have an infinitum of letters?
A direct generalization to the continuum by the introduction of a probability density, $p_a = \rho(a)da$, suffers from a breakdown of the $\mathcal Q$-interpretation: the degrees of freedom in the continuum cause the entropy to diverge.
To see this, simply rewrite Eq. \ref{ShannonInfo} as
\ee{\label{Continuum Entropy}
S&=-\lim_{da\rightarrow 0}\sum_{a\in \mathcal A}\rho(a)da\log\rho(a)da\nonumber\\
&=-\int da\rho(a)\log \rho(a)+\lim_{da\rightarrow 0} \log\frac{1}{da}.
}
The second term introduces a logarithmic divergence which spoils the continuum generalization.

Differential entropy attempts to fix this problem by simply ignoring this infinite shift, and using the first term in Eq. \ref{Continuum Entropy}
\ee{
\mathcal S = \int\! da \ \rho(a)\log\rho(a).
}
Though finite, it suffers from two ailments.
First, it is not invariant under a change of coordinates.
This is clear once one recalls that the probability density transforms as a scalar density under coordinate transformations: when $x\rightarrow\bar x$, the density transforms as $\rho(x)\rightarrow |\frac{\partial \bar x}{\partial x}|\bar\rho(\bar x)$. However,
this is not so much of a problem, given that the coordinates play the role of an alphabet.
If we explore the languistic analogy further, phonemes requiring one symbol in one alphabet may require two symbols in another: 
information is measured relative to a {\it fixed} alphabet.
Second, it is not positive definite, making a $\mathcal Q$-interpretation problematic.

Differential CE (DCE) sets out to measure the informational complexity of a particular field configuration, while alleviating the non-positivity of differential entropy.
Consider some energy density field $\rho(\bf r)$, localized in space.
By this we mean that it has a bounded $\ell^2$-norm.
The decomposition into wave modes is given by the Fourier transform:
\ee{
\tilde \rho(\mathbf k) = (2\pi)^{-\frac{d}{2}}\int\!\! d^d\mathbf r \  \rho(\mathbf r) e^{-i\mathbf k \cdot \mathbf r}.
}
A detector sensitive to the full spectrum of wave modes will detect a wave mode within a volume $d^d\mathbf k$ centered at $\mathbf k$ with probability proportional to the power in that mode:
\ee{
p(\mathbf k|d^d\mathbf k) \propto |\tilde \rho(\mathbf k)|^2d^d\mathbf k.
}
This power spectrum will be peaked at some particular scale, $|\mathbf k_*|$.
The relative contribution of a wave mode to $\mathbf k_*$ is the {\it modal fraction},
\ee{
f(\mathbf k) = \frac{p(\mathbf k|d^d\mathbf k)}{p(\mathbf k_*|d^d\mathbf k)} = \frac{|\tilde\rho(\mathbf k)|^2}{|\tilde\rho(\mathbf k_*)|^2}.
}
The DCE is defined from the modal fraction as
\ee{
\mathcal C[\rho] = -\int\!\! d^d\mathbf k \ \ f(\mathbf k)\ln f(\mathbf k).
}
Since the modal fraction is globally $\le1$, the positivity of DCE is guaranteed.
The $\mathcal C$ is clearly for {\it Configurational}, however it also serves to remind us that the expression is dependent on the {\it coordinate system} being used. 
DCE has units of nats per unit volume.

Note that the modal fraction is proportional to the power spectrum.
Since the power spectrum is the Fourier transform of the two-point correlation function, DCE is capturing scale information. We interpret DCE as an informational measure of spatial {\it complexity}.

For completeness, we review the case of spherical symmetry, where some care must be taken.
The hyper-spherical Fourier transform reads:
\ee{\label{sFT}
\tilde\rho(k) = k^{1-\frac{d}{2}}\int_0^\infty\!\!\! dr \ r^\frac{d}{2}\rho(r)J_{\frac{d}{2}-1}(kr),
}
where $J_\nu$ are Bessel functions. 
A detector sensitive to scale will measure modes with probability $|\tilde\rho(k)|^2d^d\mathbf k$.
Hence the modal fraction reads:
\ee{\label{mf}
f(k) = \frac{| \tilde\rho( k)|^2}{|\tilde\rho(k_*)|^2}.
}
The spherical DCE is then calculated as
\ee{\label{sDCE}
\mathcal C[\rho] =- \frac{2\pi^{d/2}}{\Gamma(\frac{d}{2})}\int_0^\infty dk \ k^{d-1}f(k)\log f(k).
}

\section{Configurational Entropy and Lifetime of Oscillons}

In their simplest form, oscillons are localized, time-dependent, spherically-symmetric solutions of the nonlinear Klein-Gordon equation in a double-well potential \cite{Bogolubsky, Gleiser}.
 One of their most remarkable properties is their longevity: contrary to naive expectations, which would estimate the lifetime of an unstable bubble-like field configuration of approximate radius $R$ to be $R$ (from now on, we use units of $c = \hbar = k_B = 1)$, in 3d oscillons can live for $10^{3-4} R$, and longer in $d = 2$ \cite{GleiserSornborger, Salmi}. 
 In this manuscript, we will restrict our investigation to these simplest oscillons, although our methods could be extended to treat many different types, including oscillons that emerge in models with more than one interacting field, such as other scalars or gauge fields and gravity. 
(A sample of more complex oscillons is listed in the references.) 
We are also leaving aside the fascinating possibility that certain oscillons may have an incredibly long (infinitely long?) lifetime \cite{Honda}, a point we intend to get back to in a forthcoming manuscript.

Our goal in this paper is to obtain an estimate of the oscillon lifetime using the information theoretic methods described in the previous section. 
For this, we need first to generate oscillons from an initial condition. There are many ways to do this, and we expect that in realistic settings such as phase transitions \cite{GleiserHowell, GleiserThorarinson}, or preheating cosmological scenarios \cite{GleiserGrahamStamatopoulos, AminEasther}, spherically-symmetric oscillons will emerge from general asymmetric large-amplitude scalar field fluctuations as attractor solutions in field configuration space \cite{Gleiser, GleiserSicilia}. Indeed, a previous detailed analysis has shown this to be the case for the simple oscillons we are treating here \cite{Adib}: asymmetric configurations evolve towards spherically-symmetric oscillons by rapidly radiating away their extra initial energy. (We are not considering small-amplitude scalar field oscillons, as treated in Refs. \cite{GrahamStamatopoulos, Farhi, Amin} and other works.)

Irrespective of the particular oscillon formation mechanism, our main result is that the oscillation amplitude of the time-dependent DCE of a general oscillon configuration is smaller the longer-lived the oscillon is. Moreover, we also found that the DCE is a \textit{pattern discriminator}, distinguishing initial configurations that evolve into oscillons from other short-lived nonperturbative subcritical bubbles.

\subsection{Oscillons in Brief} 

Since oscillons have been treated extensively in the literature, we present here only their essentials, following the notation and conventions of Ref. \cite{Gleiser}, which the reader can consult for more details.

The action for a real scalar field in 3 + 1 dimensions is 

\begin{equation}\label{Action}
S[\phi(\boldsymbol{x},t)] = \int d^4x \bigg[\frac{1}{2}\partial_{\mu}\phi \: \partial^{\mu}\phi - V(\phi)\bigg].
\end{equation}
To investigate oscillons in a variety of contexts, we choose an asymmetric double well potential written as 

\begin{equation}\label{potential}
V(\phi) = \frac{m^2}{2}\phi^2 - \frac{(\epsilon + 1) \sqrt{\lambda} m}{\sqrt{2}}\phi^3 + \frac{\lambda}{4}\phi^4.
\end{equation}
Here, $\epsilon$ is the asymmetry parameter: $\epsilon = 0$ gives degenerate potential minima, and increasing $\epsilon$ increases the asymmetry. Considering spherically symmetric field configurations and rescaling to dimensionless variables $r \rightarrow r/m$, $t \rightarrow t/m$, and $\phi \rightarrow m \phi / \sqrt{\lambda}$, we obtain the Klein-Gordon equation of motion to be solved,

\begin{equation}\label{KleinGordon}
\frac{\partial^2 \phi}{\partial t^2} = \frac{\partial^2 \phi}{\partial r^2} + \frac{2}{r}\frac{\partial \phi}{\partial r}  - \phi +\frac{3}{\sqrt{2}}(1+\epsilon)\phi^2-\phi^3.
\end{equation}

We choose the initial configuration to have a Gaussian profile, with core amplitude $\phi_C$ that measures the deviation from $\phi_0$, a minimum of the potential. Oscillons are usually interpreted as nonperturbative fluctuations away from the vacuum. In the case of an asymmetric potential, they would be fluctuations away from the false vacuum and into the true. (As are critical bubbles.) The only restriction on $\phi_C$ is that it must be larger than $\phi_{\rm inf}$, the inflection point of the potential, for oscillons to exist. For simplicity, we choose it to be the location of the other potential minimum:

\begin{equation}\label{GaussianInitialProfile}
\phi(r, t=0) = \left(\phi_C-\phi_0\right) e^{-(r/R_0)^2} + \phi_0,
\end{equation}
where $R_0$ is the initial radius of the Gaussian bubble. The two minima are at $\phi_0 = 0$ and $\phi_C = \frac{3(\epsilon+1)}{2\sqrt{2}} \left[1 + \sqrt{1-16/9(\epsilon+1)^2}\right].$ To solve Eq. \ref{KleinGordon}, it is necessary to introduce a set of boundary conditions:

\begin{align*}
\phi(r \rightarrow \infty, t) &= \phi_0;\\
\phi'(r = 0, t) &= 0;\\
\dot{\phi}(r, t=0) &= 0.
\end{align*}
These conditions enforce that the field approaches the proper vacuum at spatial infinity, that the field at the core is regular, and that the bubble begins its evolution from rest. A solution $\phi(r, t)$ has energy density,

\begin{equation}\label{EnergyDensity}
\rho(r,t) = \left(\frac{1}{2}\dot{\phi}^2+\frac{1}{2}\nabla\phi^2+V(\phi)\right).
\end{equation}
The spherical Fourier transform is computed from Eq. \ref{sFT},

\begin{equation}\label{FourierTransform}
\widetilde{\rho}(k,t) = \frac{4\pi}{k}\int_0^{\infty} dr \: r \: \rho(r,t) \: \sin(kr).
\end{equation}

From this we compute the modal fraction, Eq. \ref{mf}, and use it to find the DCE, Eq. \ref{sDCE},
\begin{equation}\label{DifferentialEntropy}
\mathcal{C}(t) = - 4 \pi \int_0^{\infty} dk \: k^2 \: f(k,t) \: \log\left(f(k,t)\right).
\end{equation}
In what follows, we explore the time-dependent behavior of the $\mathcal{C}(t)$, showing that its features correlate with the longevity of oscillons and their discrimination from other bubbles.

\subsection{Numerical Methods}
\label{NumericalMethods}

\begin{figure}[h!]
\includegraphics[width=\linewidth]{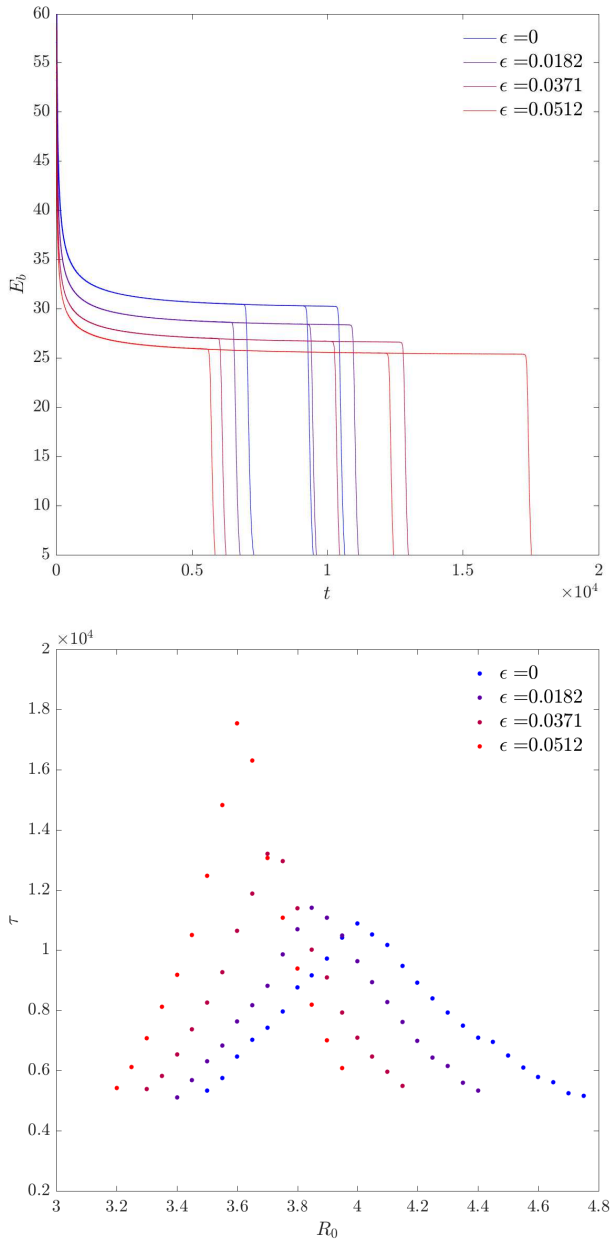}
\caption{Top: Energy $E_b$ of bubbles evolving into oscillons as a function of time for several different asymmetries $\epsilon$ (color) and $R_0$. For each asymmetry, the radius of the longest-lived oscillon is represented, along with two nearby radii. For $\epsilon = 0$ (symmetric potential), $R_0 = {3.9, 4, 4.35}$; $\epsilon = 0.0182$, $R_0 = {3.75, 3.85, 4.2}$; $\epsilon = 0.0371$, $R_0 = {3.6, 3.7, 4.05}$; and $\epsilon = 0.0512$, $R_0 = {3.5, 3.6, 3.95}$.  Potential asymmetry increases from blue to red; blue is the degenerate case. Bottom: Lifetime $\tau$ as a function of $R_0$ for different asymmetries. Color convention for $\epsilon$ is consistent throughout the paper.}
\label{energiesAndLifetimes}
\end{figure}

We used a static one-dimensional lattice of $N = 2500$ points with lattice interval $\delta r = 0.02$. The time step $\delta t$ was set to $\delta r/10$, satisfying the Courant condition: we checked that changing the time step by a factor of two in either direction didn't affect the evolution. To calculate the spatial derivatives, a fourth-order accurate centered finite difference scheme with fixed boundary conditions was employed. At the left and right boundaries of the lattice, more points were used in the forward or backward direction to keep the accuracy of the scheme fixed over the whole lattice. Time evolution was accomplished with a leapfrog scheme accurate to second order. The field, its time derivative, and the bubble's energy density were saved every thousand time steps, and the simulation was run until the energy of the bubble $E_b/m \leq 5$. This is our operative definition of the oscillon's lifetime. As can be seen in Fig. \ref{energiesAndLifetimes}, the bubble very rapidly sheds its energy when the oscillon dies, so the definition of lifetime is not very sensitive to a specific choice of energy cutoff.

In order to eliminate the artificial effects of reflection and interference of outgoing radiation from the lattice boundary, we employed adiabatic damping \cite{GleiserSornborger}. This is implemented by modifying the equation of motion by adding the term $\gamma \dot{\phi}$, where $\gamma = \left(r/L\right)^n$, $L$ is the length of the lattice, and $n$ is an integer that was changed depending on the initial bubble configuration; for the Gaussian bubbles, a good choice is $n=10$. The advantage of this approach is that, since the bubble is localized around the origin with $R_0 \ll L$, a relatively small lattice can be used, greatly reducing computational time. The damping is negligible in the region of the bubble, but if an accurate model for the outgoing radiation were desired, this approach would be inadequate. Alternatives include a large, static lattice (computationally costly), a lattice which grows at the same rate as the outgoing radiation \cite{Gleiser}, or a conformal transformation \cite{Fodor, Honda, Ikeda}.

In order to check energy conservation in the simulation, we evolved a single bubble on a very large lattice, with no adiabatic damping, for some time well into when it had settled into the oscillon phase. We could then keep track of both the energy of the bubble and the energy of the outgoing radiation. Energy was conserved to better than $1\%$ over the simulation time. 

The DCE was computed from Eq. \ref{DifferentialEntropy}, where the Fourier transform was accomplished with the matrix multiplication $\widetilde{\rho}_{\rm m} = A_{\rm mn} \rho_{\rm n}$, with the matrix $A$ given by

\begin{equation}\label{FourierTransform2}
A_{\rm mn} = \frac{4\pi\delta r}{k_{\rm m}}r_{\rm n}\sin\left(k_{\rm m} r_{\rm n}\right).
\end{equation}

We determined the energy of the bubble by integrating the energy density to a radius of $5 R_0$, which was always less than the lattice size $L$ and within the regime of negligible adiabatic damping. In the Fourier transform, then, the smallest mode that can be reliably detected (the UV cutoff) corresponds to a wavelength of $\lambda = 10 R_0$, so that $k_{\rm min} = \pi/5R_0$. The upper cutoff is taken to be $k_{\rm max} = \pi/(3 \delta r)$, a criterion equivalent to requiring that each wavelength is sampled by at least six points. We observed that this was sufficient to guarantee convergence; increasing the upper cutoff did not change the modal fraction or $\mathcal{C}$ in any measurable way.

\begin{figure}[h!]
\includegraphics[width=0.5\textwidth]{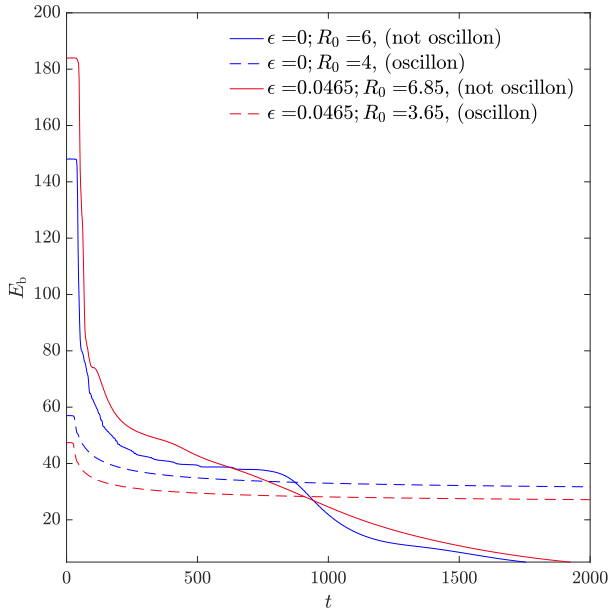}
\caption{Two examples of non-oscillon subcritical bubble energy profiles (solid lines) that erroneously triggered our oscillon criterion, from potentials with two different asymmetries. For comparison, an oscillon energy profile from the same potential is shown for each (dashed lines). Because both non-oscillons start with large initial radii, their transient radiative phase is longer, and they happen to approach the plateau energy halfway through their lives.
However, they never display the characteristic behavior of an oscillon.}
\label{energiesOscillonsAndNot}
\end{figure}

Before analyzing the DCE, we had to distinguish oscillons from other bubbles that did not form oscillons. For a given asymmetry $\epsilon$, after initially radiating away some energy, the field configurations may or not reach a well-defined plateau energy $E_p(\epsilon)$, as is described in more detail in section \ref{Results}. Even though there is a small amount of radiation, oscillons are defined as the configurations that reach $E_p$. $E_p$ was determined for each asymmetry from the longest-lived oscillon. All other bubbles were then compared to this, and if a bubble had reached $E_p$ halfway through its lifetime, it was called an oscillon. A few bubbles with larger $R_0$ happened to be at the plateau energy halfway through their evolution; these were excluded manually because none of their nearby neighbors were oscillons. Fig. \ref{energiesOscillonsAndNot} compares oscillon energy profiles to that of two examples of extraneous bubbles. 

\begin{figure}[htb!]
\includegraphics[width=\linewidth]{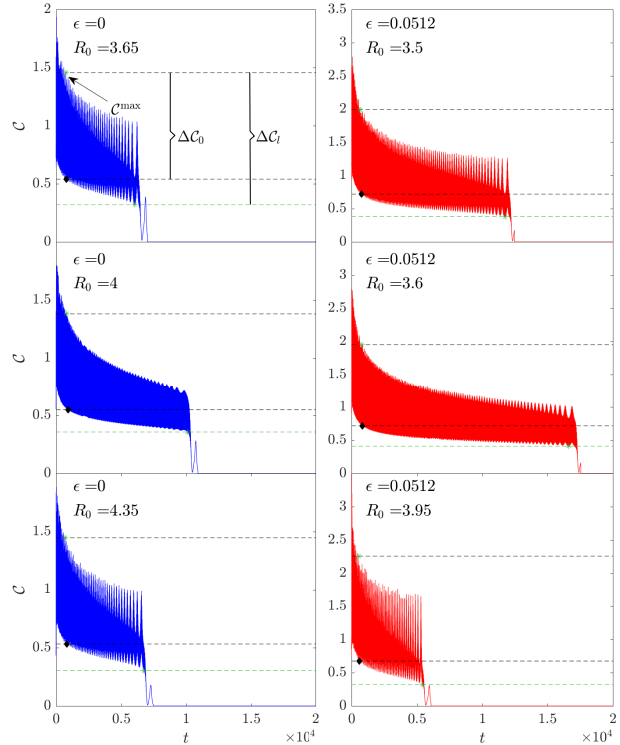}
\caption{Time evolution of $\mathcal{C}(t)$ for the symmetric potential ($\epsilon=0$, left panels) and one with a large asymmetry (right panels), and select radii. The center row is the longest-lived oscillon for our resolution. The upper (black) dashed line shows $\mathcal{C}^{\rm max}$ at the beginning of the oscillon's lifetime, so at $t_0\pm \Delta t_0$. This criterion guarantees that $\mathcal{C}^{\rm max}$ is found when $E_b$ is within ~5\% of the plateau energy. The lower (black) line shows the minimum $\mathcal{C}$ on a narrow window ($\sim t_0\pm \Delta t_0$) around the beginning of the oscillon's lifetime, $\mathcal{C}_0^{\rm min}$. The lowest (green) dashed line shows the minimum $\mathcal{C}$ over the whole lifetime of the oscillon, $\mathcal{C}_l^{\rm min}$. $\Delta\mathcal{C}_0 = \mathcal{C}^{\rm max} - \mathcal{C}_0^{\rm min}$ is the amplitude just as the oscillon has formed, while $\Delta\mathcal{C}_l = \mathcal{C}^{\rm max} - \mathcal{C}_l^{\rm min}$ is the difference between the maximum and minimum values of $\mathcal{C}$ taken from the whole lifetime of the oscillon.}
\label{sampleCEs}
\end{figure}

Because the stability of several localized physical systems has been previously associated with critical behavior of the DCE, we examined three different quantities derived from $\mathcal{C}$, described visually for a few examples in Fig. \ref{sampleCEs}. Here, we offer a brief description of how these quantities and their associated uncertainties were determined; they will be discussed in more detail in later sections. $\mathcal{C}^{\rm max}$ is the maximum value of $\mathcal{C}$ at a time $t_0$ after the initial radiative stage is complete, and the oscillon stage is reached. $t_0$ is determined by considering when $\Delta E_b \lessapprox 0.01 E_0$, where $\Delta E_b$ is defined between two sample times. Typically, we chose $\Delta t = 0.05$, sufficiently smaller than any time scale in the system. 
$\Delta\mathcal{C}_0$ is the amplitude of the oscillations in $\mathcal{C}$ around time $t_0$, defined as $\Delta\mathcal{C}_0 = \mathcal{C}^{\rm max} - \mathcal{C}_0^{\rm min}$, where $\mathcal{C}_0^{\rm min}$ is determined over a short window around $t_0$, $\Delta t_0 = 40,000 \delta t$. At the other end of an oscillon's existence, it very rapidly sheds its remaining energy. Just before the oscillon enters this second (and final) rapid radiative phase, we can find the minimum value of $\mathcal{C}\equiv \mathcal{C}_l^{\rm min}$, and define another amplitude of $\mathcal{C}$ oscillations over the whole \textit{lifetime} of the oscillon: $\Delta\mathcal{C}_l = \mathcal{C}^{\rm max} - \mathcal{C}_l^{\rm min}$. The time $t_l$ at which the oscillon enters its second radiative phase is determined by considering when $\Delta E_b \gtrapprox 0.01 E_0$: $t_l\lesssim \tau$, the oscillon's lifetime.

To estimate the uncertainty in $\mathcal{C}^{\rm max}$, we found the maxima in the ranges $(t_0 - 2\Delta t_0)$ and $(t_0 + 2\Delta t_0)$, denoted $\mathcal{C}_{-}^{\rm max}$ and $\mathcal{C}_{+}^{\rm max}$, respectively. The uncertainty is

\begin{equation}\label{uncertainty}
\sigma^{\rm max} = \mathcal{C}^{\rm max} - \frac{1}{2}\left(\mathcal{C}_{-}^{\rm max}+\mathcal{C}_{+}^{\rm max}\right).
\end{equation}
Uncertainties in the minima were estimated analogously, and uncertainties in $\Delta\mathcal{C}_0$ and $\Delta\mathcal{C}_l$ were determined through standard error propagation.

\subsection{Results}
\label{Results}
The general behavior for bubbles with initially Gaussian profiles that evolve into oscillons is shown for different potential asymmetries (color) and initial radii in Fig. \ref{energiesAndLifetimes}. A bubble's excess initial energy is quickly radiated away, and the oscillon forms around $t \sim 500 m^{-1}$. During the oscillon stage, the energy remains nearly constant at the plateau energy $E_p$. For the symmetric potential case, $\epsilon = 0$, $E_p \sim 32 m/\lambda$, and the plateau energy decreases as the asymmetry increases.
Oscillon lifetimes are very sensitive to the initial radius, as seen on  Fig. \ref{energiesAndLifetimes} (bottom) and, in finer resolution, in Ref. \cite{Honda}. Since we are interested here in establishing a general relationship between the differential entropy and lifetimes of oscillons, we don't need to go into such fine resolutions, which are very CPU-intensive. Work along these lines is currently in progress. Given that our approach depends on the functional profile of $\mathcal{C}$, it should be valid for any initial radius.

\begin{figure}[h!]
\includegraphics[width=\linewidth]{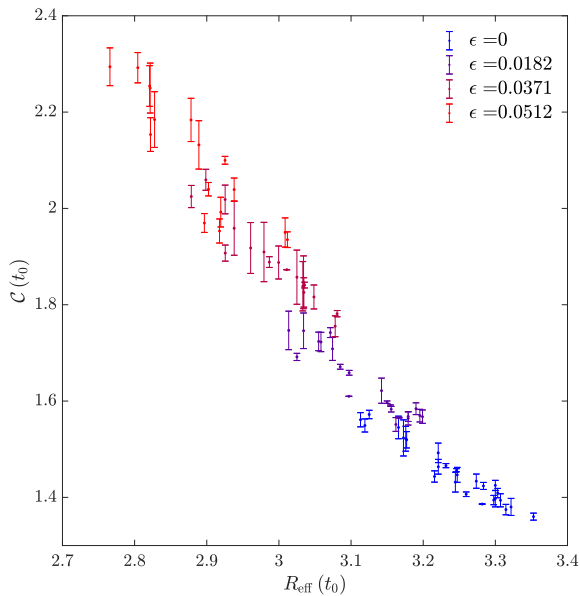}
\caption{The relationship between the differential configurational entropy $\mathcal{C}$ and the effective radius $R_{\rm eff}$ at the beginning of the oscillon stage (time $t_0$). The effective radius is generally smaller for larger asymmetries, and $\mathcal{C}$ is larger for smaller $R_{\rm eff}$.}
\label{CmaxWithReff}
\end{figure}

From Ref. \cite{GleiserStamatopoulos1}, the DCE of a 3d Gaussian with radius $R_0$ is $\mathcal{C} = \frac{3}{2}\left(2\pi/R_0^2\right)^{3/2}$. The more localized the configuration in space, the higher its DCE. While the oscillon's field profile does not remain Gaussian at later times, this relationship between the size of a system and its entropy is general. Fig. \ref{CmaxWithReff} supports this assertion, showing $\mathcal{C}(t_0)$ as a function of effective radius $R_{\rm eff}$, at time $t_0$ when the bubble had just entered the oscillon stage.  $R_{\rm eff}$ is computed as the first moment of the bubble's energy density.

\begin{figure}[htb!]
\includegraphics[width=\linewidth]{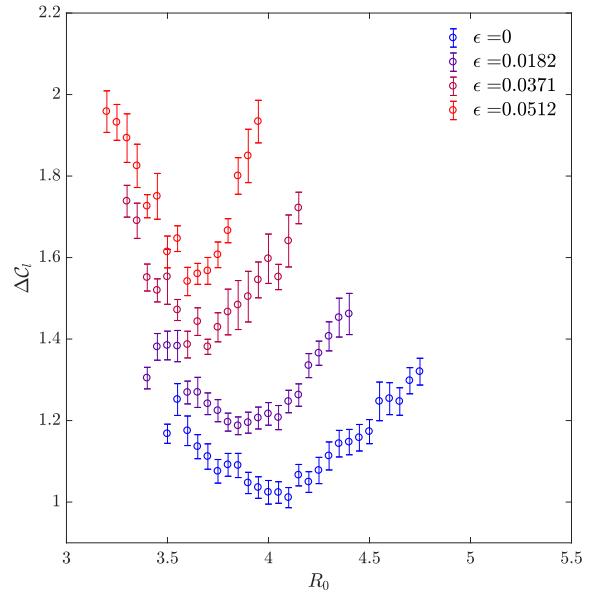}
\caption{$\Delta\mathcal{C}_l$ and corresponding uncertainties are shown as a function of initial radius for different asymmetries. Note that, in each case, the longest-lived oscillon has the smallest $\Delta\mathcal{C}_l$. See numerical methods section for a description of the uncertainty estimation.}
\label{sampleCEs2}
\end{figure}

Back to Fig. \ref{sampleCEs}, we see some examples of $\mathcal{C}(t)$ for two different asymmetries and radii. A few features are apparent. Qualitatively, the evolution of $\mathcal{C}(t)$ is similar for all cases. However, the quantitative differences are essential. First, $\mathcal{C}(t)$ oscillates about a mean as a function of time, with amplitudes that vary with the lifetime. The longest-lived oscillons have the smallest amplitudes. Second, the frequency of oscillations in $\mathcal{C}(t)$ decreases as the oscillon approaches the end of its life, which correlates with other studies for the field's core amplitude $\phi(0,t)$ \cite{Gleiser, GleiserSicilia}. In Fig. \ref{sampleCEs2}, we explore the first point in more detail, showing $\Delta C_l$ for different initial radius $R_0$ for several asymmetries. In all cases, the longest-lived oscillons are at the minimum of $\Delta C_l$.

From these results, we tried to determine a measure that correlated with the lifetime of oscillons. Previous works for static solitonic \cite{GleiserSowinski1} and stellar \cite{GleiserJiang1} configurations have established that a maximum in DCE relates to the instability point of a given configuration. Following this lead, but now for a time-dependent system, we defined the three measures introduced in Section \ref{NumericalMethods}, 
$\mathcal{C}^{\rm max}$, $\Delta\mathcal{C}_0 = \mathcal{C}^{\rm max} - \mathcal{C}^{\rm min}_0$ (near the beginning of the oscillon stage), and $\Delta\mathcal{C}_l = \mathcal{C}^{\rm max} - \mathcal{C}^{\rm min}_l$ (over the whole lifetime of the oscillon). 

We then examined how well these three measures correlated with the oscillon's lifetime. $\mathcal{C}^{\rm max}$ and $\Delta\mathcal{C}_0$ have the advantage of occurring near the beginning of the bubble's evolution, and $\mathcal{C}^{\rm max}$ is the time-dependent extension of previous work (\cite{GleiserSowinski1,GleiserJiang1} and others) in which the maximum DCE for a system coincides with the onset of instability. However, given that we also know that longer-lived oscillons are known to be better virialized \cite{Gleiser}, we might expect that those which have the smallest amplitude in DCE over the whole course of oscillon evolution are the longest-lived.

\begin{figure*}[htb!]
\includegraphics[width=\textwidth]{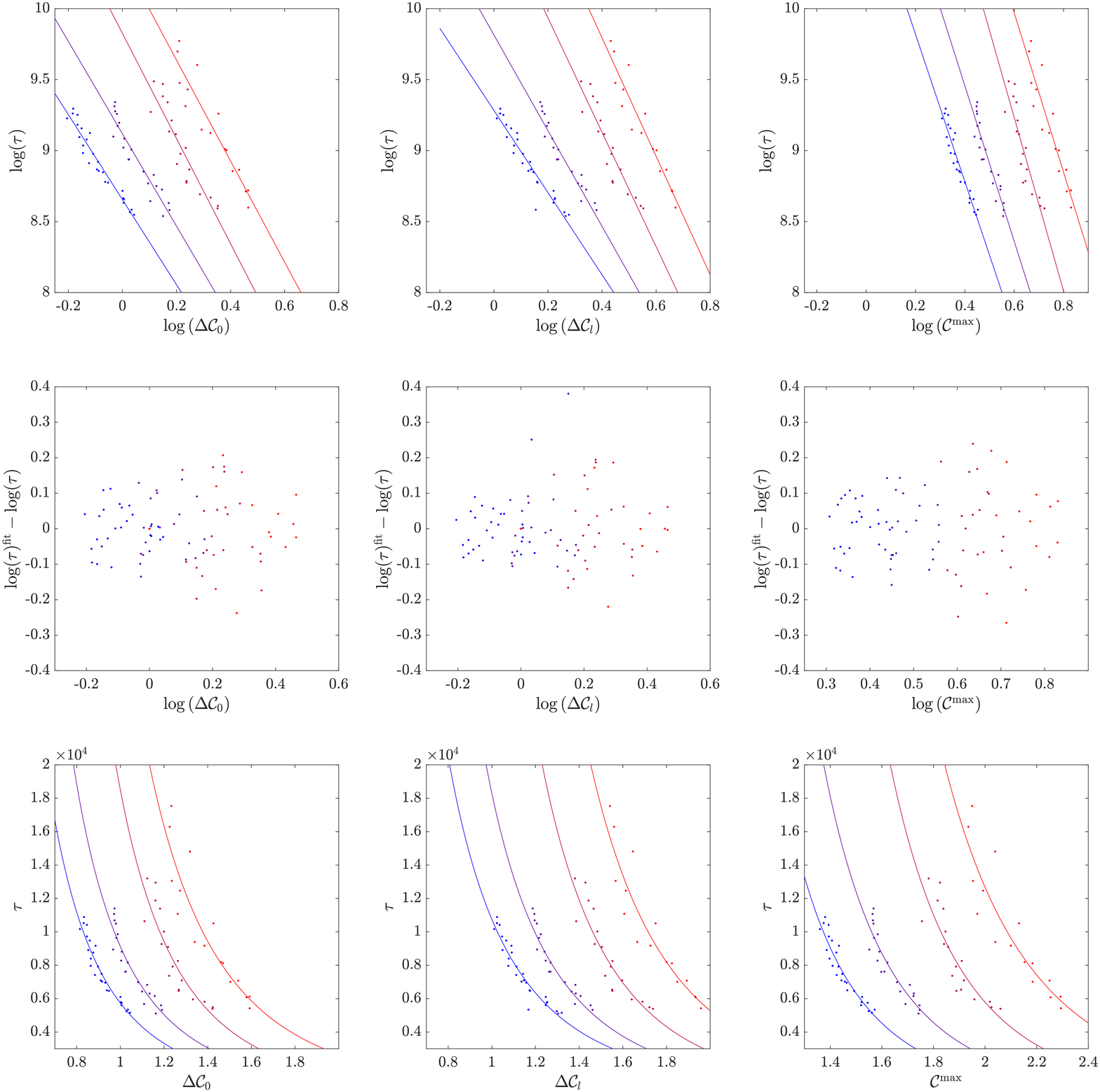}
\caption{From left to right, the columns show the trends between oscillon lifetime and $\Delta\mathcal{C}_0$, $\Delta\mathcal{C}_l$, and $\mathcal{C}^{\rm max}$, respectively (see Fig. \ref{sampleCEs} for a graphical definition of these quantities). The top row shows the linear fits to the log-log plots, while the bottom row shows the power law relationships constructed from these fits. The center row shows the residuals.}
\label{fitsToOtherCEs}
\end{figure*}

\begin{figure}[htb!]
\includegraphics[width=\linewidth]{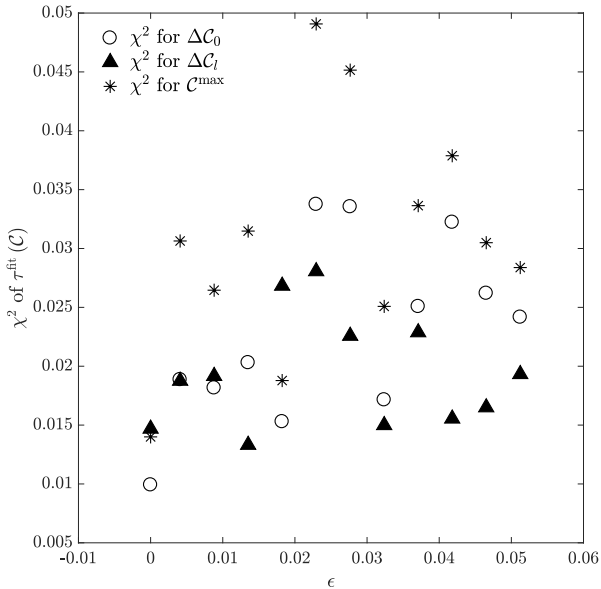}
\caption{$\chi$-square statistics for fits from Fig. \ref{fitsToOtherCEs} and other values of $\epsilon$. Note that for the degenerate potential, the best fit is for $\Delta\mathcal{C}_0$, while $\mathcal{C}^{\rm max}$ and  $\Delta\mathcal{C}_l$ give practically identical results. For nine of the twelve asymmetries, $\Delta\mathcal{C}_l$ has the lowest $\chi$-square, and thus offers the best overall fit to the lifetime.}
\label{chiSquare}
\end{figure}

In Fig. \ref{fitsToOtherCEs} we show the results for the three DCE measures for four different potential asymmetries. The bottom row of Fig. \ref{fitsToOtherCEs} shows the power law relationships reconstructed from the log-log fits of lifetime for each measure. Each of these is reasonable, and none is clearly better than the other, especially for small asymmetry. Indeed, for the degenerate potential, $\epsilon=0$, the results are very similar, although computing the $\chi$-square, shown in Fig. \ref{chiSquare}, we observe that $\Delta\mathcal{C}_l$ offers a slightly better fit ($<0.027$) as the asymmetry increases.

Given that the $\chi$-squares for $\Delta\mathcal{C}_0$ and for $\mathcal{C}^{\rm max}$ are not much larger (both below $\chi^2 < 0.05$), we could substitute either one for $\Delta\mathcal{C}_l$ throughout this paper and observe the same trends. The choice depends on the specific needs of the investigation: if we were interested in a fast estimate of the oscillon  lifetime without the need for great accuracy, both $\Delta\mathcal{C}_0$ and $\mathcal{C}^{\rm max}$ offer a distinct computational advantage, since the oscillon would not need to be simulated until reaching its ultimate fate. These results establish the DCE as a {\it predictor} of oscillon lifetime with reasonable precision. (See below.) For higher precision, one would use $\Delta\mathcal{C}_l$, although this would not be advantageous from a computational perspective, only for investigating the relationship between DCE and oscillons over their entire lifetime.

\begin{figure}[htb!]
\includegraphics[width=\linewidth]{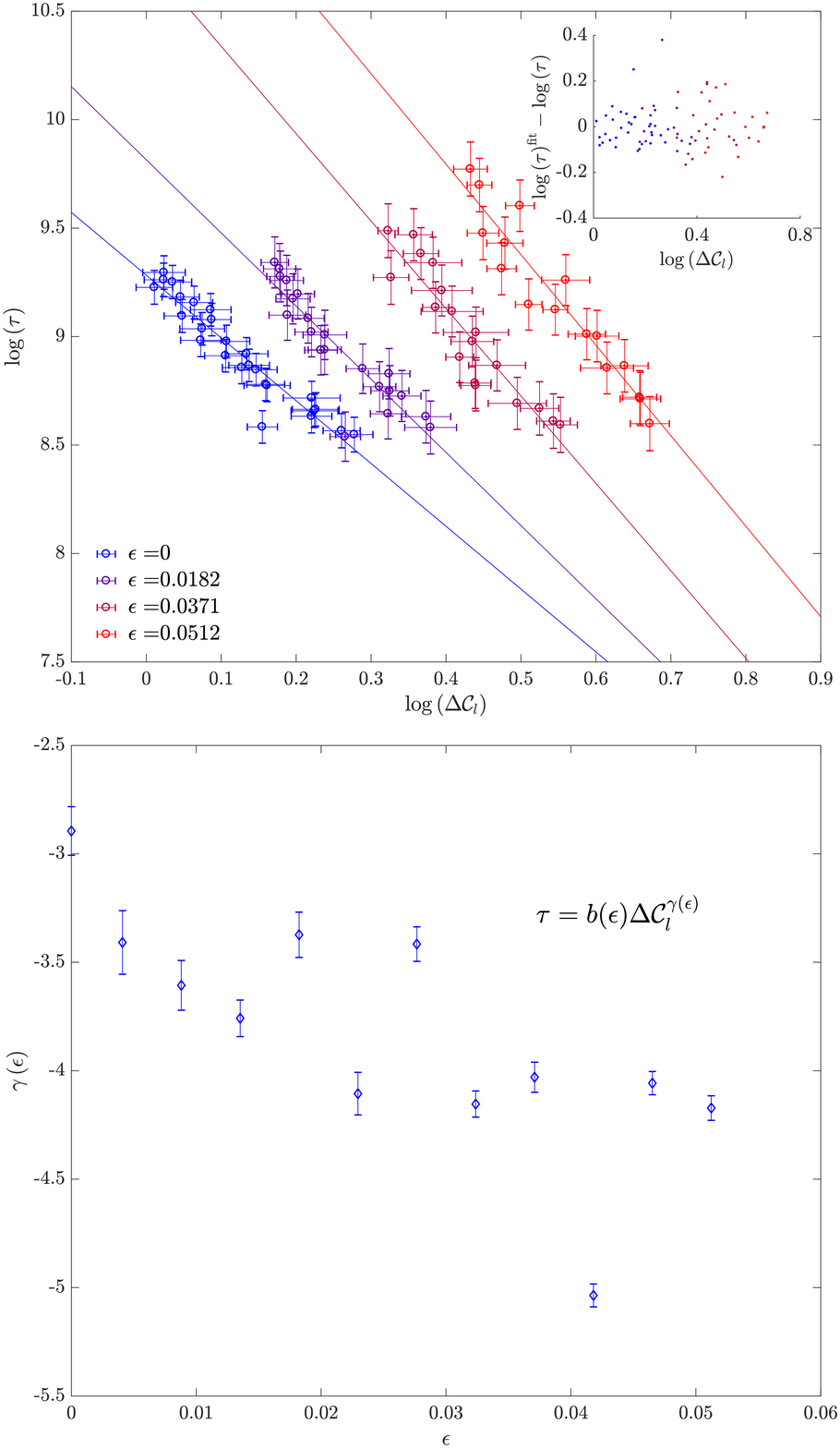}
\caption{Top: Lifetime is plotted against $\Delta\mathcal{C}_l$ on a log scale, and a linear fit to the data for each asymmetry is shown. The trend shows clearly that longer-lived oscillons have smaller DCE amplitudes in all cases, and that DCE can be used as a predictor of oscillon lifetime with reasonable precision. (More below.) The residuals for the fits are shown in the insert . Assuming the lifetimes are independent and normally distributed, the error estimate on the lifetimes includes half of future predictions that would be measured. Bottom: The slopes of the linear fits for several asymmetries along with their uncertainty estimates.}
\label{fitsToLogLifetime}
\end{figure}

In Fig. \ref{fitsToLogLifetime}, we show the log-log fits to lifetime $\tau$ versus $\Delta\mathcal{C}_l$, along with the residuals (insert). We extract a scaling relationship between $\tau$ and $\Delta\mathcal{C}_l$, which we write as

\begin{equation}\label{scalingLaw}
\tau = b(\epsilon)\Delta\mathcal{C}_l^{\gamma(\epsilon)}.
\end{equation}
The asymmetry-dependent scaling exponent $\gamma$ is shown with $\epsilon$ in the insert of Fig. \ref{fitsToLogLifetime}. The accuracy in the exponent can be improved by saving snapshots of the field at smaller time intervals.

\begin{figure}[htb!]
\includegraphics[width=\linewidth]{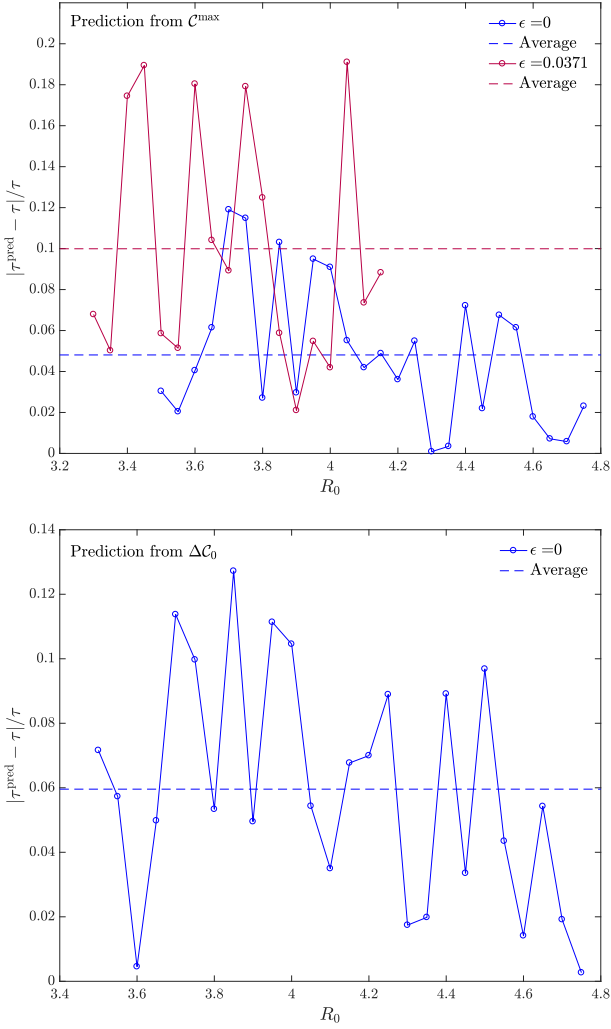}
\caption{Top: Fractional difference between the lifetime and the prediction for lifetime from $\mathcal{C}^{\rm max}$ is shown for two different asymmetries. Bottom: Fractional difference between the lifetime and the prediction for lifetime from $\Delta\mathcal{C}_{0}$ is shown, only for the symmetric potential.}
\label{lifetimePrediction}
\end{figure}

To test how useful the different DCE measures are as predictors of oscillon lifetime, in Fig. \ref{lifetimePrediction} we compare the oscillon lifetime prediction  for different initial radii reconstructed from the fits using ${\cal C}^{\rm max}$ (top) and $\Delta\mathcal{C}_0$ (bottom), with the numerical results. The averages are also shown (dashed lines). For both measures, the predictions have an average precision of under $\sim 6\%$ for $\epsilon=0$, becoming worse for larger asymmetries.

\begin{figure}[htb!]
\includegraphics[width=\linewidth]{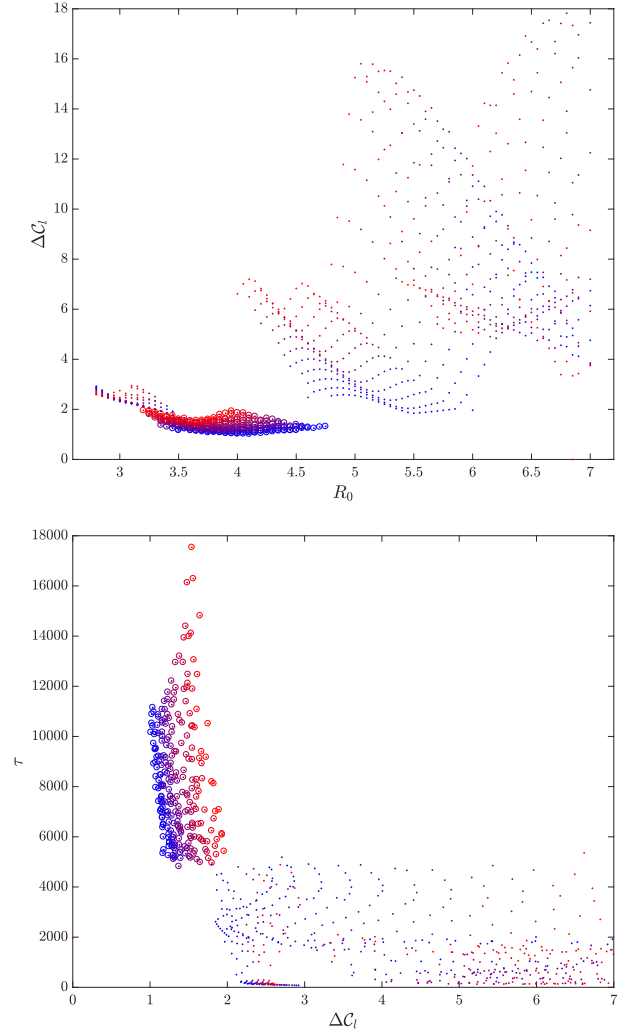}
\caption{Top: $\Delta\mathcal{C}_l$ is shown for each initial radius and asymmetry (color). We simulated several sub-critical bubbles that did not form oscillons. Those bubbles that do form oscillons are circled (lower left); they exist in a distinct region of parameter space. Bottom: Lifetime as a function of $\Delta\mathcal{C}_l$ is shown for all bubbles. Again, the oscillons (circled) are distinguished from other subcritical bubbles.}
\label{patternDiscriminator} 
\end{figure}

Our final results explore the sensitivity of the DCE as a pattern discriminator. Thus far, our results have included only those bubbles which become oscillons. However, it is well-known that some initial configurations are either too small or too large to form oscillons. Computing the DCE for those objects, we found a clear signature that distinguishes them from the initial bubbles that do evolve into oscillons. This allows us to use DCE to select the initial shapes that evolve into oscillons. In Fig. \ref{patternDiscriminator} we show on the top $\Delta\mathcal{C}_l$ versus $R_0$, and on the bottom $\tau$ versus $\Delta\mathcal{C}_l$, for a large sample of initial Gaussian bubbles. Each bubble that forms an oscillon is circled; the space occupied by the oscillons is clearly distinct. The DCE naturally divides the bubbles into two disjoint classes: those that evolve into oscillons and those that don't. Similar results can be obtained using $C^{\rm max}$ and $\Delta C_0$ with appropriate selection criteria.

\section{Concluding Remarks}

In this work, we applied methods from information theory, in particular, from a measure of spatial complexity known as differential configurational entropy (DCE) \cite{GleiserStamatopoulos2}, to the dynamics of oscillons \cite{Bogolubsky, Gleiser}. Previous results for static matter configurations, from solitonic to stellar, have determined that the DCE correlates directly with the stability of the object: maximum DCE denotes a point of instability in parameter space. Examples include a dissociating soliton or a collapsing star. We have extended these previous approaches for static systems to the time-dependent dynamics of oscillons with the main goal of using DCE to determine the longevity of oscillons, one of the key open questions in the related literature.  

To obtain an oscillon one usually starts with a localized field configuration and evolve it dynamically solving the corresponding equation of motion. It is well-known that oscillons have a shape memory, in that their longevity is determined by the details of the initial configuration: the shape of their progenitor seals their destiny. Given that DCE describes the spatial complexity of a given field configuration in terms of the Fourier transform of its two-point correlation function, we constructed three measures from DCE to investigate whether they could predict an oscillon's longevity. Our results show that all three measures can be used to predict with different levels of accuracy the oscillon's longevity based on the parameter(s) determining the shape of its initial progenitor, in our case the initial bubble radius $R_0$ and $\epsilon$, the asymmetry in the potential. In particular, two measures that use DCE at the onset of the oscillon's stage can predict the lifetime to average accuracies better than $6\%$ or so, although the accuracy worsens as the asymmetry of the potential is increased: shape indeed seals destiny. This is very useful if one wants to explore the longevity of oscillons for different parameters without committing the considerable CPU resources to run long simulations. A third measure using DCE produces better accuracy, but must be computed along the oscillon's entire lifecycle.

We also explored the use of DCE as a pattern discriminator. Investigating configurations that become long-lived oscillons and those that collapse before they reach the oscillon stage, we showed that DCE neatly distinguishes the two: unstable, short-lived bubbles tend to have higher DCE and thus occupy a different region in DCE space. 

In a forthcoming companion paper, we apply similar techniques to the study of quantum vacuum decay in several spatial dimensions, correlating DCE with false vacuum decay rates. It would be interesting to extend the current approach to investigate the longevity of oscillons in different spatial dimensions, as well as for one-dimensional breathers \cite{Campbell}. Another avenue of research is to use DCE to explore previous results that indicate that certain oscillons in $3d$ may live for indefinitely long times \cite{Honda}. Although these configurations require highly accurate numerics to be discovered and may thus be mostly of academic interest, they could shed light on the physical mechanisms that generate oscillons in the first place and their extension to gravitational theories \cite{Ikeda}. Work along these lines is currently in progress.

\acknowledgments MG and MS were supported in part by a Department of Energy grant DE-SC0010386. DS was supported by the Institute for Cross-Disciplinary Engagement through a grant from the John Templeton Foundation.


\end{document}